# Formation of Domains within Lower-to-higher Symmetry Structural Transition in CrI$_3$

P. Doležal[1], M. Kratochvílová[1], D. Hovančík[1], V. Holý[1], V. Sechovský[1], J. Pospíšil[1]

*1 Charles University, Faculty of Mathematics and Physics, Department of Condensed Matter Physics, Ke Karlovu 5, 121 16 Prague 2, Czech Republic*[2]

**ABSTRACT**

CrI$_3$ represents one of the most important van der Waals systems on the route to understanding two-dimensional magnetic phenomena. Being arranged in a specific layered structure it also provides a unique opportunity to investigate structural transformations in dimension-confined systems. CrI$_3$ is dimorphic and possesses a higher symmetry low-temperature phase, which is quite uncommon. It contrasts with vanadium trihalides which show a higher symmetry high-temperature. An explanation of this distinct behavior together with a large cycle-dependent transition hysteresis is still an open question. Our low-temperature X-ray diffraction study conducted on CrI$_3$ single crystals complemented by magnetization and specific heat measurements was focused mainly on specific features of the structural transition during cooling. Our results manifest that the structural transition during cooling relates to the formation of structural domains despite the lower symmetry structure transforming to a higher symmetry one. We propose that these domains could control the transition temperature and also the size of thermal hysteresis.

**INTRODUCTION**

The CrI$_3$ compound belongs to magnetic van der Waals (vdW) materials which have been intensively studied in recent years mainly for possible applications in spintronics and optoelectronics[1-3] but also as objects suitable for testing two-dimensional (2D) magnetic toy models[4]. The 2D character of transition metal trihalides $TX_3$ ($T$-transition metal, $X$ – halide) allows the preparation of thin layers needed for engineering applications. Chromium trihalides CrI$_3$, CrBr$_3$, and CrCl$_3$ are among the pioneering materials in 2D magnetism research. CrI$_3$ and CrBr$_3$ are ferromagnetic (FM) with $T_c$ = 61 K[5, 6] and 37 K[7], respectively, whereas CrCl$_3$ shows antiferromagnetic (AFM) order below $T_N$ = 17 K[8, 9]. Monolayers of all three compounds were reported to be FM[10-12]. The vanadium counterparts are FM (VI$_3$[13, 14]) and AFM (VBr$_3$[15] and VCl$_3$[16]).

The Cr and V trihalides have two crystal structures in common, the rhombohedral of BiI$_3$ type (space group $R$-3) and the AlCl$_3$-type monoclinic ($C2/m$)[13, 14]. The high temperature (HT) structure of the three V$X_3$ compounds is the rhombohedral one which undergoes a structural transition to the monoclinic phase with cooling through the transition temperature $T_s$ (which lies between 79 and 100 K for all three V$X_3$ compounds). On the contrary, the HT phase of Cr$X_3$ compounds is monoclinic and cooling leads to a structural transition connected with increasing the symmetry from monoclinic to rhombohedral[17, 18] which is unusual in solid state physics. This structural transformation exhibits large thermal hysteresis (more than 40 K), which is in strong contrast with the negligible hysteresis of structural transitions in VI$_3$ and VBr$_3$[16, 19]. Moreover, the size of transition hysteresis in CrI$_3$ is thermal history-dependent, in particular, it depends on the number of cooling-warming cycles over the transition as reported by McGuire *et al.*[5, 17]. Interestingly, Niu *et al.* have recently discovered that stacking of vdW layers on the surface of CrI$_3$ flakes in the low temperature (LT) phase corresponds to the bulk monoclinic structure of the HT phase and the rest of the inner layers keeps the bulk stacking order typical for the rhombohedral structure[20]. A recent synchrotron X-ray powder diffraction study on bulk $CrI_3$ samples revealed the coexistence of both monoclinic and rhombohedral crystal structures down to the 10 K[21].

The phase of the LT structure is usually less symmetric, which implies the formation of domains in the sample and consequently affects the lattice parameters. However, the reversed symmetry order of the LT and HT phases observed in $CrI_3$ raises a fundamental thermodynamic question regarding domain formation, the type of domains, and the transition mechanism itself. It is of high importance to understand this process as the magnetic properties are closely related to the stacking order of the vdW layers especially in $CrI_3$[22]. The theoretical studies have shown that the monoclinic stacking in $CrI_3$ favors an antiferromagnetic ground state, whereas the rhombohedral stacking results in ferromagnetic order[20, 22]. These findings point to the importance of studying structural transformation for understanding magnetic properties.

In this work, we studied the peculiarities of structural transition in $CrI_3$ by X-ray single crystal diffraction, magnetization and specific–heat measurements. The study revealed the formation of unusual structural domains during the transition within cooling and evidenced the relation to magnetic and thermal properties.

**EXPERIMENTAL METHODS**

The $CrI_3$ single crystals were grown by the chemical vapor transport method from the stoichiometric ratio of elements in a two-zone horizontal furnace in a temperature gradient of 650/550 °C in a sealed quartz tube for two weeks. The black reflective single crystals in the form of flat flakes of several millimeter sizes were received. The low-temperature X-ray diffraction was performed on the refurbished Siemens D500 $\theta$-$\theta$ diffractometer in the Bragg-Brentano geometry using the $Cu_{K\alpha1,2}$ radiation. The He closed-cycle was used for cooling. The temperature was controlled by the low-temperature cryostat (*ColdEdge*), with temperature stabilization better than 0.1 K and with absolute uncertainty of 0.5 K. The sample chamber was filled with He gas to ensure good thermal contact between the sample and the cold finger. A piezo-driven rotator was used for an alignment of the sample in the $\phi$-direction. This experimental setup allows measurements in the temperature range of 3 – 300 K. The reciprocal space maps were measured by position-sensitive detector Mythen 1K. The limitations in sample alignment allow the measurement of only one diffraction maximum per temperature cycle. The $CrI_3$ samples are partialy sensitive to the moisture and this is also the reason, why different single crystalline samples from the same batch were used for the specific heat, magnetization and diffraction measurements. The presented X-ray diffraction study consists of two different single crystalline sample. The first one used for the determination of lattice parameters and for study of behavior of domains during cooling. The second one was used for the study of temeperature hysteresis using the (0 0 24) maps. The samples were plates with size around 1 x 0.5 mm in both cases and whole sample surface was irradiated by X-ray beam.The magnetization curves were measured by an MPMS-7 SQUID magnetometer (*Quantum Design, Inc.*). The heat capacity data were measured by PPMS9T (*Quantum Design, Inc.*) using the heat-pulse method as the standard relaxation method principle is not sensitive to first-order phase transitions.

**RESULTS & DISCUSSION**

To understand the difference between monoclinic and rhombohedral stacking we compare the LT and HT structure in more detail. The unit cells of both crystal structures are shown in Fig. 1 (for simplicity only the Cr atoms are shown). Note that for the description of the rhombohedral structure model, we use the conventional hexagonal unit (e.g. the label $c_{hex}$ refers to the $c$ lattice parameter of the hexagonal unit cell in the rhombohedral phase). The monoclinic structure contains only one Cr layer per unit cell. On the other hand, the rhombohedral structure has three such layers resulting in $c_{hex} \sim 3c_{mono}$. The Cr atoms are ordered in hexagons within the layers. The hexagons are regular in rhombohedral structure and slightly distorted in the monoclinic phase in which each layer is offset relative to the adjacent Cr layers. This shift is the most pronounced difference between the HT and LT structures (see the colored hexagons in Fig. 1). In the monoclinic phase the layers are shifted in the direction of the hexagon vertex while in the rhombohedral structure in the direction perpendicular to the side of the hexagon. The vdW interaction between the layers is weak in comparison to the covalent bonds between Cr and I atoms within the layers, which probably control the crystal structure. One Cr atom is surrounded by 6 neighboring I atoms creating the octahedron. In both crystal structures the octahedron is not regular (see corners of Fig. 1.) In the rhombohedral structure it consists of two different equilateral triangles, while in the monoclinic crystal structure, the triangles are scalene. A direct comparison of the crystal structures is not straightforward because no t-group-subgroup relation allows the description of a more symmetrical structure in the t-subgroup. But some similarities can be inferred; $c_{hex} \approx 3c_{mono}, a_{hex} \approx a_{mono}, b_{mono} \approx \mathrm{sqr}(3)a_{hex}$ giving us a hint of changes during the structural transition. The diffraction maxima (0 0 $l$) in a symmetrical direction should be only shifted and the appearance of new ones is not expected because of extinction rules in the rhombohedral space group. This behavior was confirmed by the measurement of a symmetrical $\theta$-$2\theta$ scan (see Fig. S1 in Supporting Information[23]). Since the structural transition from a lower symmetry to a higher symmetry crystal structure, no $2\theta$-splitting of diffraction maxima connected with the creation of domains by twinning mechanism should be observed. Our results also show that the full width at half maximum of diffraction $2\theta$ profiles is comparable in both phases and slightly bigger for the rhombohedral phase (see Fig. S2[23]).

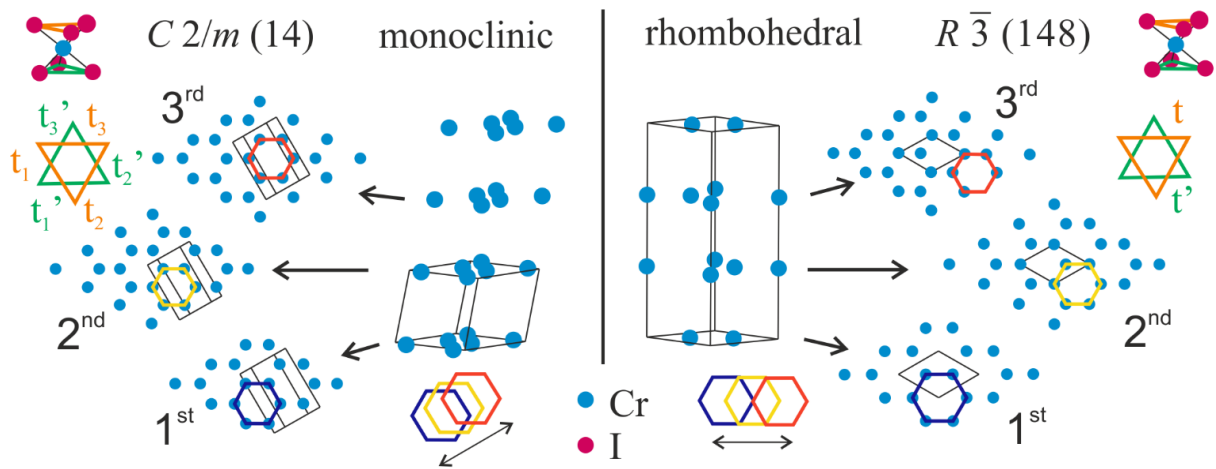

**Figure 1.** Comparison of monoclinic and rhombohedral crystal structures of $CrI_3$. The colored triangles represent the distortion of a regular octahedron. The colored hexagons show the different shifts of layers in the monoclinic and rhombohedral crystal structures. It has to be noted that the 4th layer overlaps exactly with the 1st layer (perpendicular view on the basal plane) in the rhombohedral structure. On the other hand, in the monoclinic structure, there is no such direct overlap, but the shift between the 1st and 4th layers is very small.

The reciprocal space maps of (0 0 8)$_{mono}$, (0 6 6)$_{mono}$, (-4 0 6)$_{mono}$ and (4 0 6)$_{mono}$ diffraction peaks were measured to determine the lattice parameters as a function of temperature. The maps were integrated in rocking direction to obtain $2\theta$ profiles which were fited by pseudo Voigt function. The lattice parameters were determined from these $2\theta$ positions. It is important to emphasize that given a plate-like sample's shape ($c^*_{mono}$ is perpendicular to this plate) limits a number of diffraction maxima which are accessible for measuring in our configuration (for more details about alignment see Section II). The above-listed monoclinic diffraction maxima cannot be related to the rhombohedral ones detected below the transition temperature, however, close to their positions in $2\theta$ we measured diffraction peaks $(0\ 0\ 24)_{hex}$, $(0\ 3\ 18)_{hex}$, $(-4\ 2\ 15)_{hex}/(4\ -2\ 15)_{hex}$, $(4\ -2\ 21)_{hex}/(-4\ 2\ 21)_{hex}$. The resulting lattice parameters are shown in Fig. 2. The dominant thermal contraction is observed for $c_{mono}$ and $c_{hex}$ lattice parameters. The ferromagnetic ordering at 61 K relates to the change in the slope of $c_{hex}$ lattice parameter. No such anomaly is observed in the basal plane in $a_{hex}$ lattice parameter. The $\beta$-angle in the monoclinic structure exhibits only a small change of 0.012° during cooling. To compare the volumes of monoclinic and rhombohedral phases, the molar volume is plotted in Fig. 3a) (comparing the volume of unit cells would be misleading since there is no group subgroup relation). During the transition, the volume reduces by almost 0.5%, which is quite a huge change in comparison to the vanadium-based vdW compounds like $VBr_3$ (0.08 %)[16] and $VI_3$[19] where the change of volume was negligible. The change of the volume during the ferromagnetic transition is on the border of experimental error.

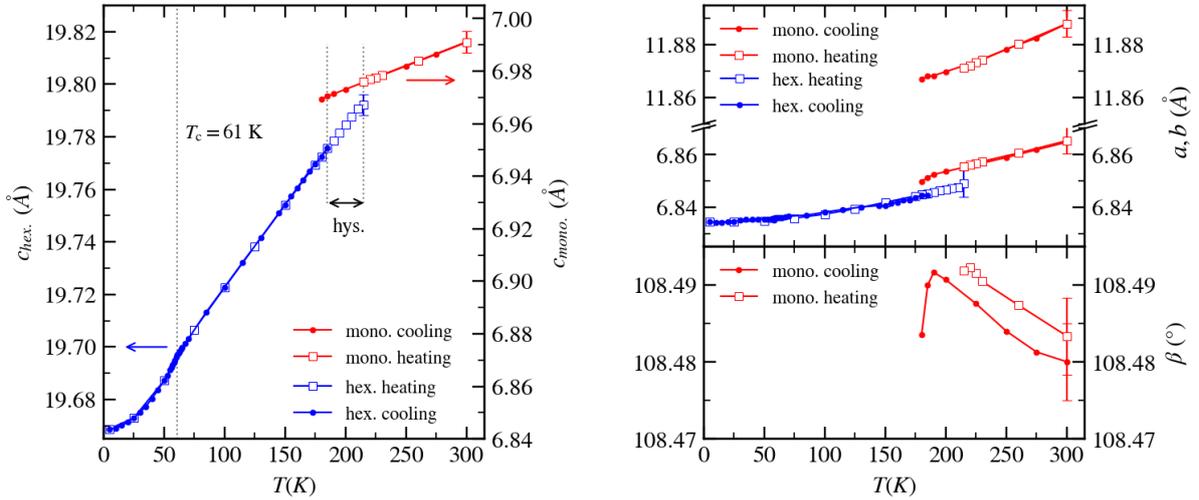

**Figure 2**. Temperature dependence of the lattice parameters in monoclinic and rhombohedral crystal structures. The hexagonal unit cell is used for the rhombohedral crystal structure and the corresponding lattice parameters are labelled by subscript hex. The experimental error connected with the precission is smaller than the size of the data point. The error bars in the plots show the error conneted with the accuracy.

The vdW gap is directly related to the distance between Cr layers $d_{Cr-Cr}$ and not to the $c$ lattice parameter since in the monoclinic structure $c$ is not perpendicular to the **ab** plane. The $d_{Cr-Cr}$ dependence is shown in Fig. 3 b). In both crystal structures, the vdW gap decreases with decreasing temperature having a higher slope in the rhombohedral phase. All diffraction maxima were measured during cooling and heating to test whether the change of lattice parameters is reversible. The overlap of heating and cooling curves (see full and open symbols in Fig. 2 and Fig. 3 and Fig. S3[23]) demonstrate no effect of temperature cycling.

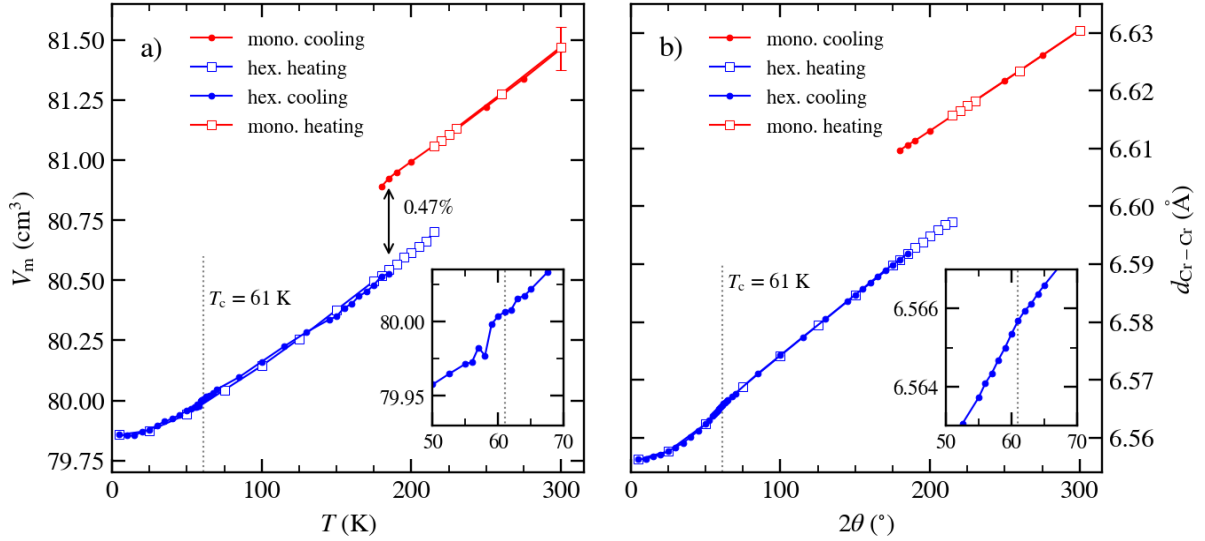

**Figure 3.** a) Molar volume of monoclinic and hexagonal crystal structures as a function of temperature. The inset shows the behavior close to the ferromagnetic transition. b) Temperature dependence of distance between Cr layers, which is proportional to the vdW gap. The experimental error connected with the precission is smaller than the size of the data point. The error bar in the a) panel shows the error conneted with the accuracy.

A large temperature interval of phase coexistence during the structural transition was already observed in previous studies, where the dependence of its size on thermal cycles was reported[5]. In our experimental setup, only one diffraction maxima can be measured during the cooling and heating cycle. Therefore, we tested the thermal robustness of the hysteresis by choosing $(0\ 0\ 8)_{mono}$ and $(0\ 0\ 24)_{hex}$ diffraction peaks measured on a single crystalline sample without a cooling history. We used the integral intensity of the peaks to determine the ratio of the monoclinic and rhombohedral phases within the coexistence interval. Fig. 4a), b) displays the monoclinic volume fraction as a function of temperature for five temperature cycles. It has to be noted, that the diffraction maps of $(0\ 0\ 8)_{mono}$ contains two separated mosaic blocks refered as a part 1 (P1) and part 2 (P2). Each part has a quite large mosaicity up to 2.5° (see Fig. S4, S5 [23]). Despite having identical lattice parameters, P1 and P2 exhibit entirely different hysteresis behavior, see Fig. 4a), b). During cooling the transition in P1 happens in two steps. The fractional volume of 40 % transform to the rhombohedral phase at ≈ 175 K and the rest of 60 % at a substantially lower temperature of ≈ 117 K. During heating the transition has a character of a sharp step at 215 K. The behavior of P2 during cooling is different, because the lowest transition happens around 165 K. This phenomenon is understandable from the diffraction map, where can be easily found that also in P1 each transition is connected with different domain in the sample (see Fig. S5[23]). The P1 is a sum of mosaic blocks with large (≈ 96 K) and small (≈ 40 K) hysteresis. As Fig. 4a) shows the cooling/heating cycling gradually shrinks the large hysteresis of P1 in an asymmetric way i.e. by shifting the lower transition temperature up during cooling while the transition upon heating remains almost intact. After the fifth cycle, the large and small hysteresis in P1 are almost identical and comparable to the hysteresis of P2. Similar evolution, depending on temperature cycling displays the magnetization curve in Figure 4c).

This observation in previous paragraph raises a question of interpretation of the difference between domains with large and small hysteresis. The chemical composition can be excluded as the lattice parameters are identical for both domains. Presumably, mechanical properties like grain size, the number of defects, etc. might be responsible for the effect. However, that would imply that the transition is connected with the formation of defects or new domains in the rhombohedral phase. Especially the creation of new domain groups would be quite unconventional assuming that the transition is from a structure of lower symmetry to one of higher symmetry structure with no twinning as is usually the case of a symmetrically reversal transition.

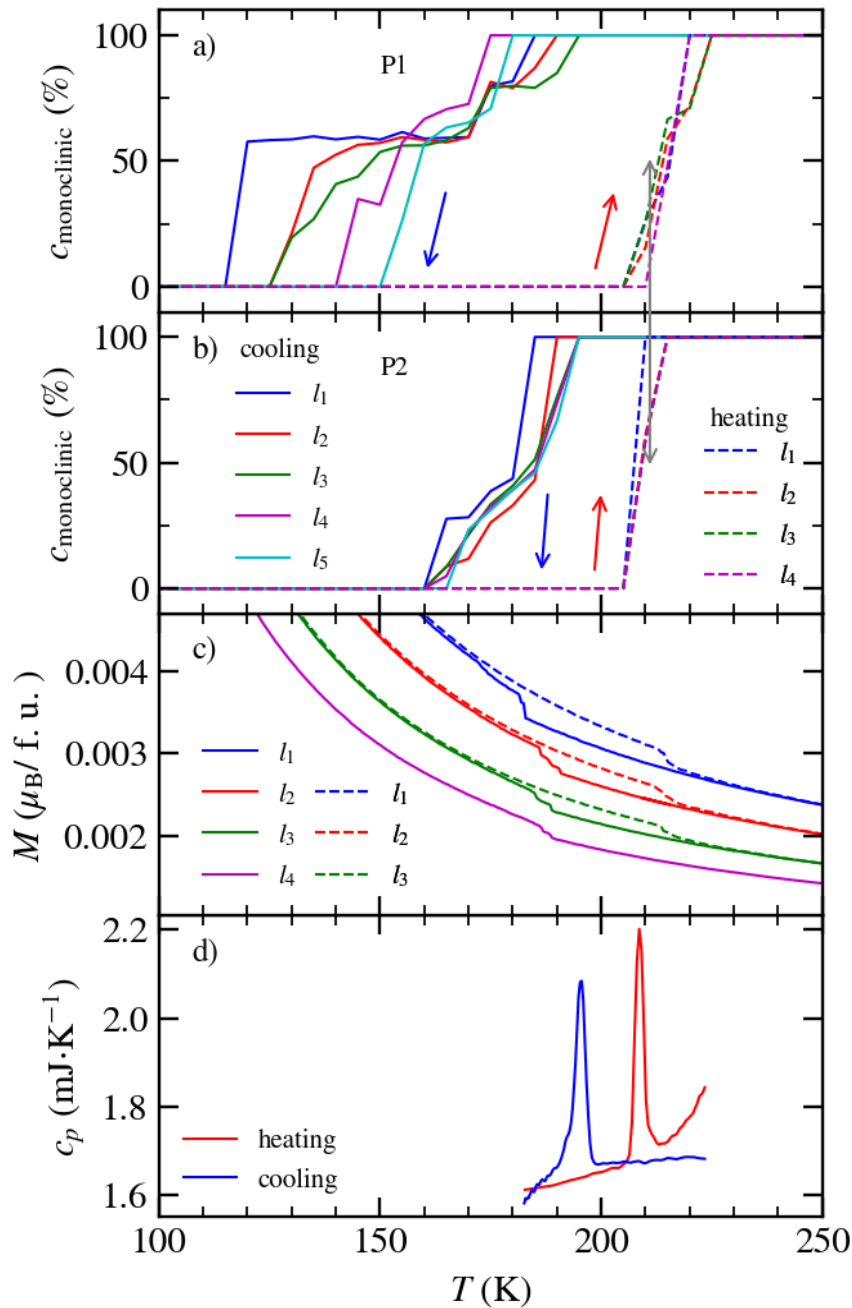

**Figure 4.** a) and b) Concentration of monoclinic phase in the sample as a function of temperature. The $l_i$ refers to the number of temperature cycles (one cycle represent the cooling and heating curve). P1 and P2 show the behavior of contributions having different $\omega_{offset}$. b) A change of hysteresis in magnetization curves during the structural transition. d) Specific heat of $CrI_3$ single crystal measured by single pulse method. The samples used for diffraction, magnetization and specific heat measurement were different.

To get a deeper picture of what happens with the domains during the structural transition, we investigated the $\omega_{offest}$ profiles of (-4 0 6)$_{mono}$ and (-4 2 15)$_{hex}$ diffraction peaks. The $\omega_{offest}$ is an angle between diffraction vector and normal vector of the sample surface (see Fig. S6[23]). The (-4 0 6)$_{mono}$ diffraction peak is in $a_{mono}c_{mono}$ plane (see Fig. S7[23]). We emphasize that $\omega_{offest}$ is very sensitive to the sample's tilt which is changing during cooling and heating. Therefore, the direct comparison of peak shapes and their intensities could be misleading, and only qualitative information about the behavior of the whole $\omega_{offset}$ profile is relevant. In the monoclinic phase, we can see a shift of the whole $\omega_{offset}$ profile shown in Fig. 5, which is a result mainly of lattice parameter change and tilt of the sample during cooling. In the rhombohedral phase right after the phase transition, the $\omega_{offset}$ profile for (-4 2 15)$_{hex}$ comprises two main components. These components start to separate from each other during further cooling (see Fig. 5).

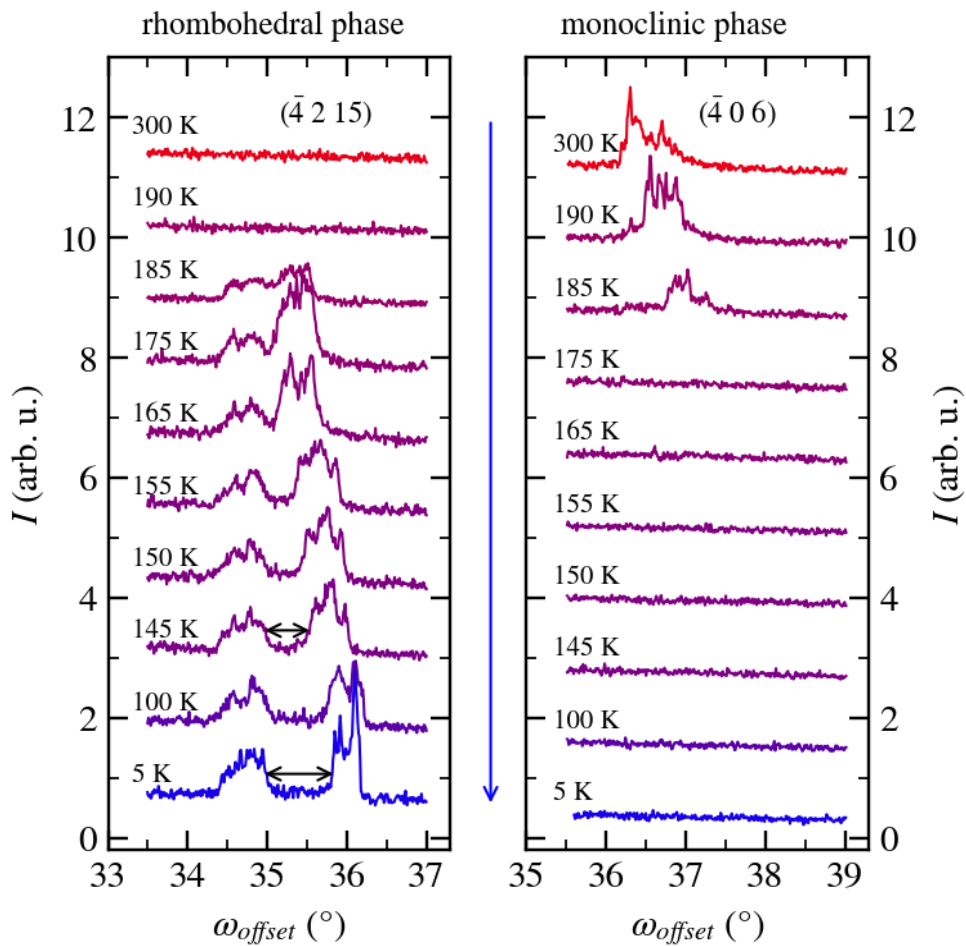

**Figure 5.** $\omega_{offset}$ profiles of (-4 2 15)$_{hex}$ and (-4 0 6)$_{mono}$ diffraction maxima. Two different contributions and their separation during cooling are visible in the first panel. The blue arrow indicates the cooling cycle. The curves are shifted along the *y*-direction for better readability.

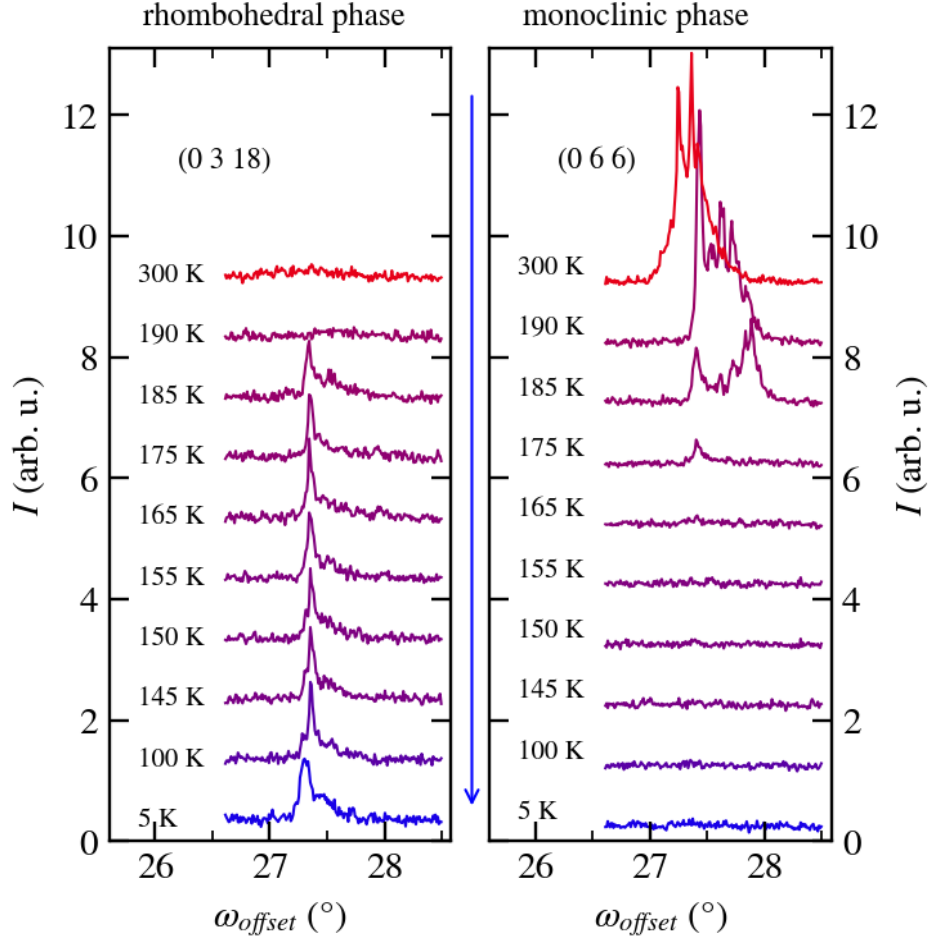

**Figure 6.** $\omega_{offset}$ profiles of $(0\ 3\ 18)_{hex}$ and $(0\ 6\ 6)_{mono}$ diffraction maxima. Only one contribution is visible in the rhombohedral and monoclinic phases during cooling. The blue arrow indicates the cooling cycle. The curves are shifted along the *y*-direction for better readability.

On the other hand, no such splitting was observed in the case of $(0\ 6\ 6)_{mono}/(0\ 3\ 18)_{hex}$ diffraction, see Fig. 6. Fig. 7 shows the temperature dependencies of these $\omega_{offset}$ profiles. One component of $(-4\ 2\ 15)_{hex}$ is almost temperature independent, while the second component is rapidly increasing its $\omega_{offset}$ between 180 K and 115 K and saturates below 100 K. This observation corroborates a scenario of the domain's structure re/formation during the transition and their thermally induced movement. During heating the movement of the second component is opposite returning to the starting $\omega_{offset}$ position, see Fig. 7. Interestingly, the transition to the HT phase takes place exactly at the temperature where the $\omega_{offset}$ separation of the components is approximately the same as at the transition temperature during cooling. This demonstrates a direct impact of domains on the hysteresis of structural transition and explains the possibility of having domains in the sample with different hysteresis behavior as we observed in samples without cooling history (see Fig. 4).

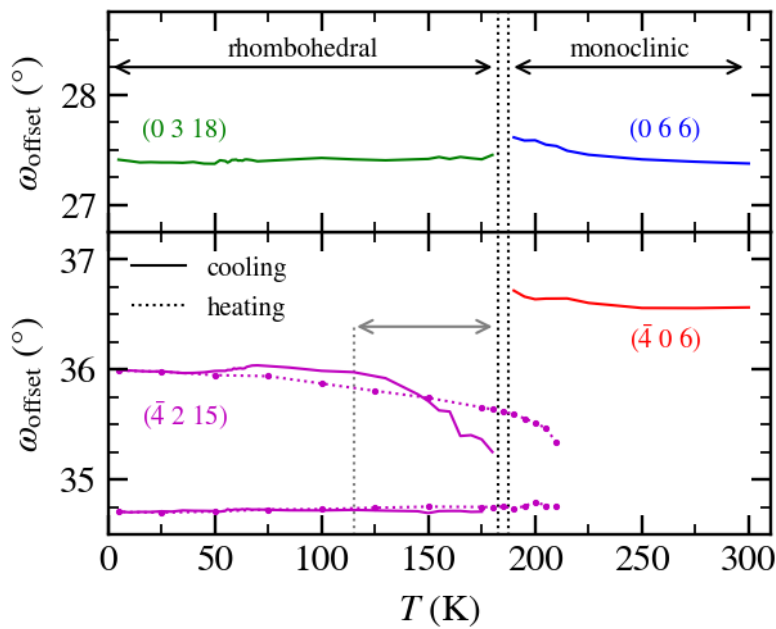

**Figure 7.** Comparison of the temperature dependence of $\omega_{\text{offset}}$ profile of $(0\ 6\ 6)_{\text{mono}}$, $(-4\ 0\ 6)_{\text{mono}}$, $(-4\ 2\ 15)_{\text{hex}}$ and $(0\ 3\ 18)_{\text{hex}}$ diffraction maxima. The grey arrow indicates a temperature interval of 115 – 185 K with the biggest movement of domains. The narrow interval between dotted black lines corresponds to the coexistence of both phases during the transition at the cooling cycle.

To point out the differences, we compare the structural transition in CrI$_3$ with the V$X_3$. In these compounds, the LT phase has lower symmetry than the HT phase[13, 16, 19]. Therefore, the structural transition results in the formation of domains, however, these domains have different origins i.e. originate from a lattice distortion. Hence, the temperature dependence of lattice parameters $a$ and $b$ is opposite, similar to a transition from austenite to martensite[24]. In contrast, the domains in CrI$_3$ are formed without distortion of the lattice. As a result, the structural transition is much more sensitive to the real structure of the sample and its mosaicity, whereas, in VI$_3$ and VBr$_3$ the domain structure is established in each mosaic block separately, hence, the transition is less sensitive to the original domain/mosaic structure. Our results can also explain, why the monoclinic and rhombohedral phases coexist in CrI$_3$ down to 10 K in the powder X-ray diffraction study reported by Maseguer-Sánchez et al.[21]. In their data, the transition is broadened and at 10 K 10 % of the sample is still in the monoclinic phase. The structural transition then looks unfinished. The explanation might originate in the mechanical treatment to get a powder sample since as follows from our observation the transition is very sensitive to structural defects and grain boundaries.

CrCl$_3$ exhibits very similar behavior showing cycle-dependent hysteresis and having comparable temperatures of structural transition[18] suggesting that a similar mechanism of structural transition like in CrI$_3$ can be expected. In CrBr$_3$ the structural transition has not been studied in detail. Only the transition temperature 423 K is known from the literature[25].

## CONCLUSIONS

We studied the structural transition in the van der Waals compound CrI$_3$ by X-ray single crystal diffraction, magnetization and specific–heat measurements and discussed the results in comparison with vanadium trihalides. We determined the lattice parameters as a function of temperature in range 3-300 K and specifiy the change of the volume during structural transition.

Our study revealed the formation of new domains groups within cooling $CrI_3$ when the lower symmetry (monoclinic) structure transforms to a more symmetric rhombohedral structure. In this case the domains cannot form due to a distortion of the crystal lattice within the transition. During cooling also the transition temperature strongly depends on the thermal history of the sample (115 – 185 K), whereas the transition temperature during heating remains intact (around 215 K) by any change in thermal history. It seems that the domain structure together with lattice defects are the main control parameters of the transition temperature and the hysteresis.

**ACKNOWLEDGMENTS**

This work is a part of the research project GAČR 21-06083S which is financed by the Czech Science Foundation and project GAUK 938220 financed by the Charles University Grant Agency. The single-crystal growth and characterization, and experiments in steady magnetic fields were carried out in the Materials Growth and Measurement Laboratory MGML (see: http://mgml.eu) which is supported within the program of Czech Research Infrastructures (project no. LM2018096). This project was also supported by OP VVV project MATFUN under Grant No. CZ.02.1.01/0.0/0.0/15_003/0000487.

Supporting Information for:

# Formation of domains within lower-to-higher symmetry structural transition in CrI$_3$

P. Doležal[1], M. Kratochvílová[1], D. Hovančík[1], V. Holý[1], V. Sechovský[1], J. Pospíšil[1]

[1]*Charles University, Faculty of Mathematics and Physics, Department of Condensed Matter Physics, Ke Karlovu 5, 121 16 Prague 2, Czech Republic*

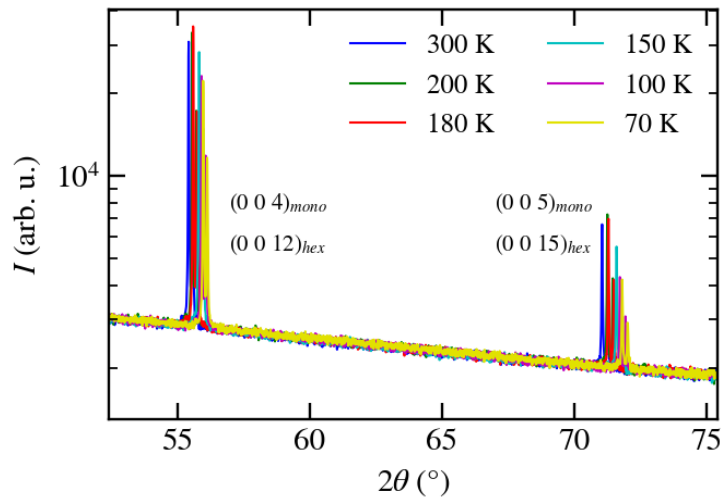

Fig. S1: $\theta - 2\theta$ scan between (0 0 12)$_{hex}$ and (0 0 15)$_{hex}$ diffraction maxima showing that there is no other intensity between them, which would be a sign of disorder in $c_{hex}$ direction. It has to be note that the Cu$_{K\alpha1,2}$ radiation was used and the diffraction peaks are doublets.

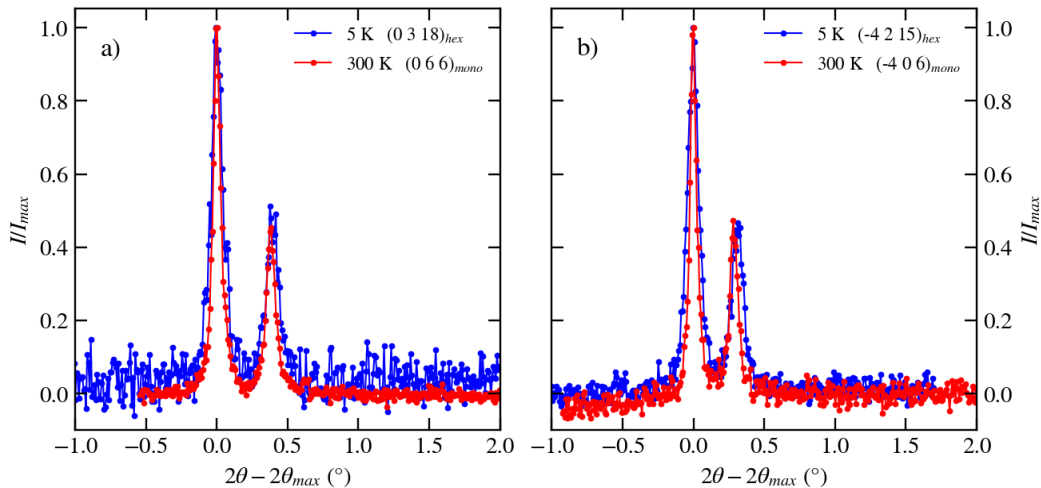

Fig. S2: Comparison of $2\theta$ profiles in the monoclinic and rhombohedral phases. The hexagonal and monoclinic peaks are on almost the same $2\theta$ angle and therefore they could be compared. In general slightly bigger full width at half maximum is observed for the rhombohedral phase. It has to be note that the Cu$_{K\alpha1,2}$ radiation was used and therefore in panelas a) and b) the diffraction peaks are doublets.

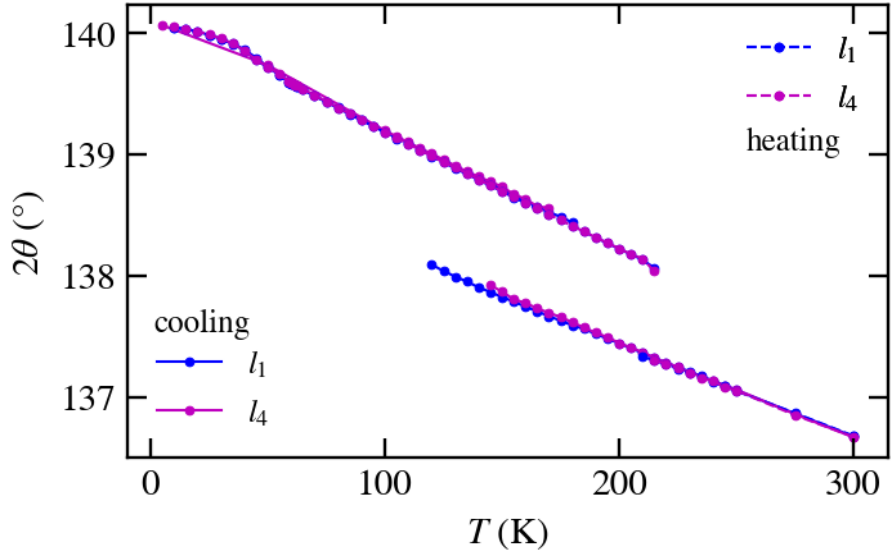

Fig. S3: Temperature dependence of $(0\,0\,24)_{hex}$ diffraction maxima. The overlap of $l_1$ and $l_4$ data points shows the independence of lattice parameters on temperature cycles.

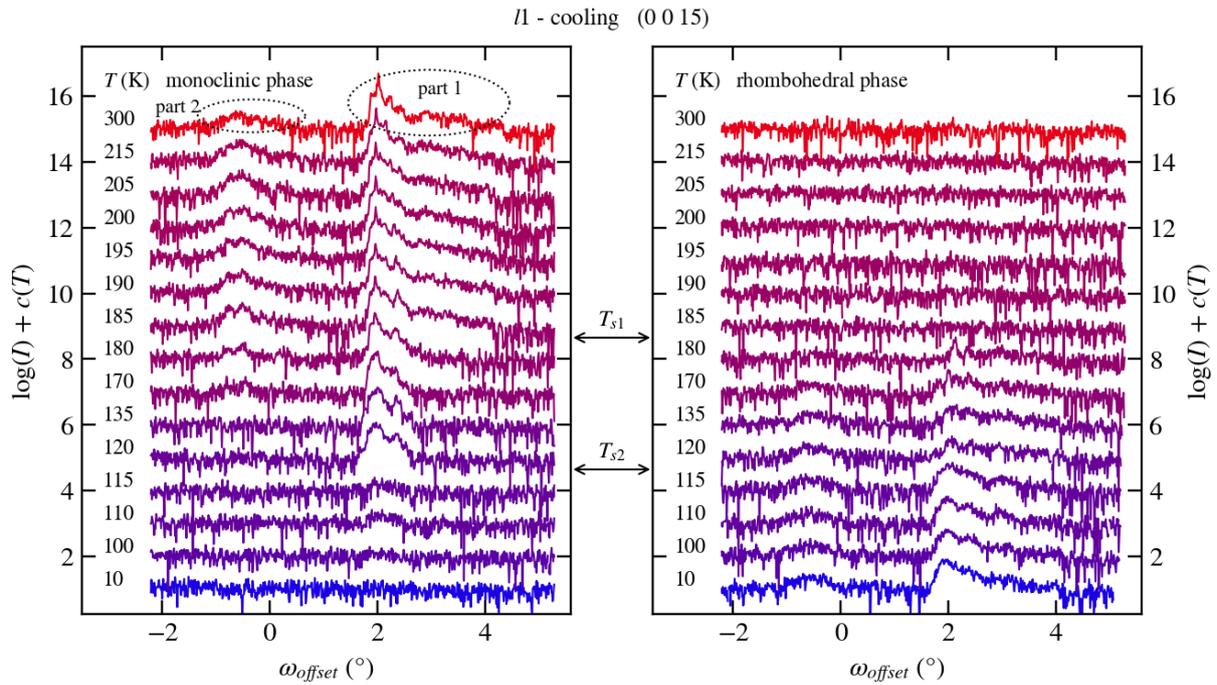

Fig. S4: $\omega_{offset}$ profile of $(0\,0\,15)_{hex}$ diffraction maximum. It is possible to observe two distinct parts (part 1, part 2) and their distinct transition temperatures $T_{s1}$ and $T_{s2}$. For more details see the main text.

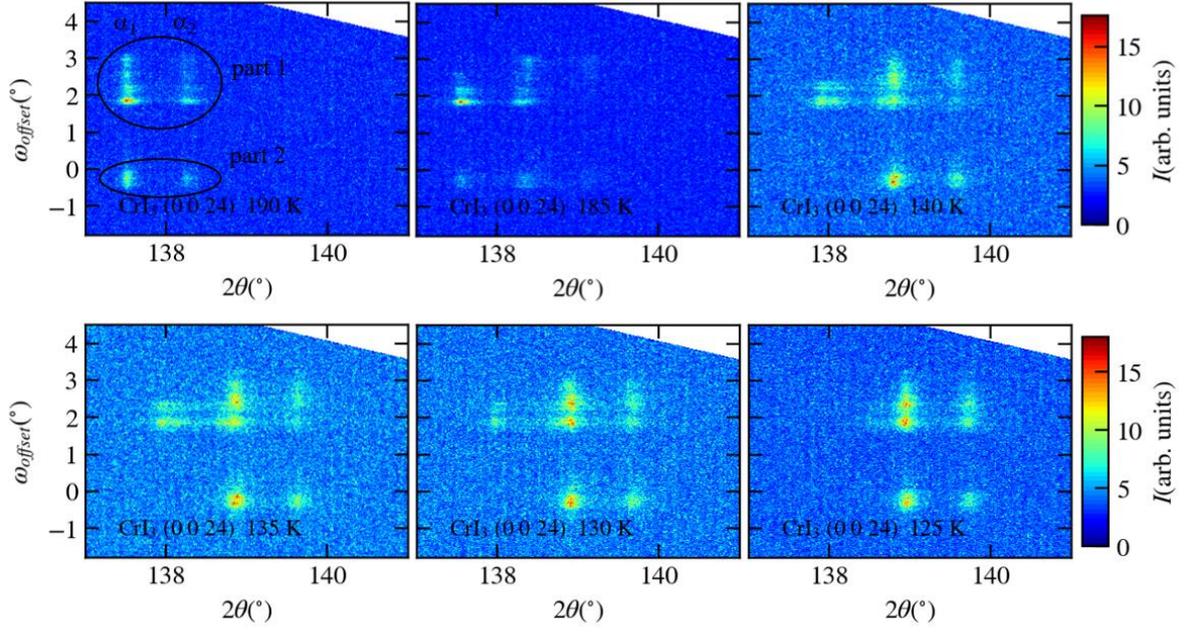

Fig. S5: $\omega_{offset}$-$2\theta$ maps of (0 0 24)$_{hex}$ diffraction maximum showing the temperature dependence through the structural transition in CrI$_3$ during cooling $l_2$. It has to be note that the Cu$K\alpha_{1,2}$ radiation was used and the diffraction peaks are doublets, see labels $\alpha_1$ and $\alpha_2$.

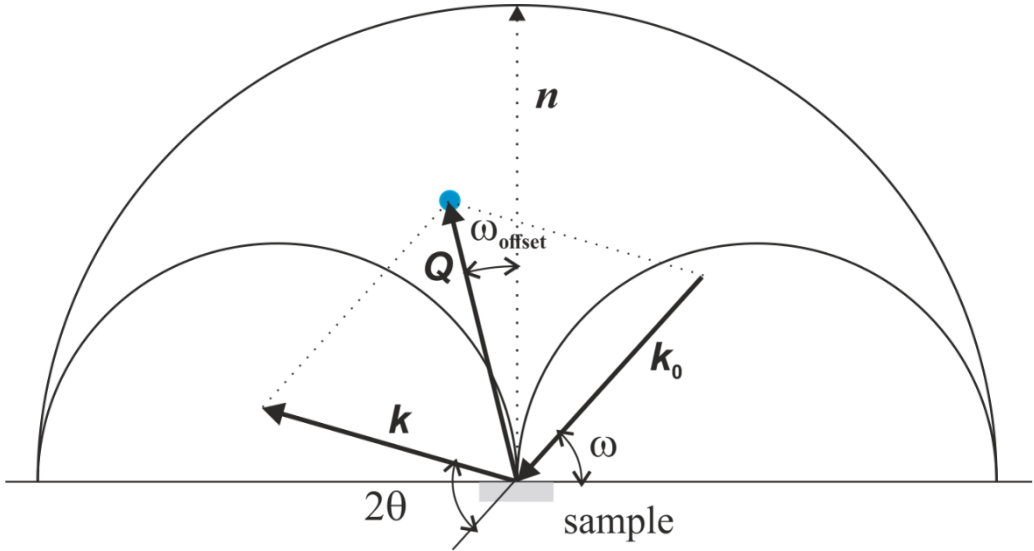

Fig. S6: Graphical representation of angles $\omega$, $2\theta$ and $\omega_{offset}$ corresponding to angle of incidence, angle between incident and diffracted beam and angle between diffraction vector $Q$ and normal $n$ of the sample surface respectively. The $k_0$ and $k$ are wave vectors of incident and diffracted beam.

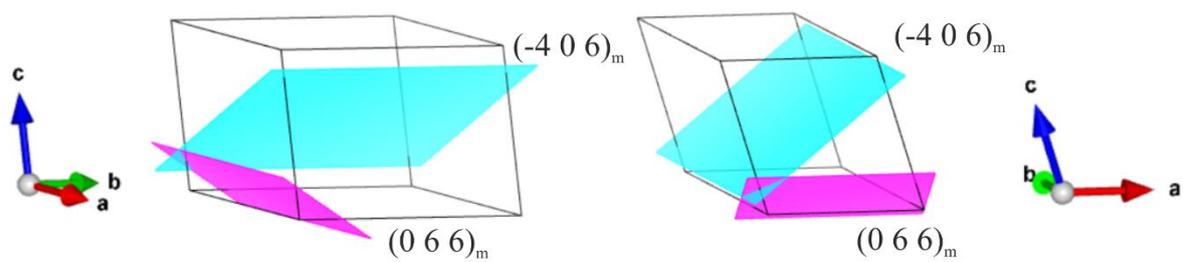

Fig. S7: Orientation of (0 6 6)$_m$ and (-4 0 6)$_m$ lattice planes within the monoclinic unit cell. The figures were plot using the VESTA software [1].

**REFERENCES**

1. K. Momma and F. Izumi, VESTA 3 for three-dimensional visualization of crystal, volumetric and morphology data, J. Appl. Crystallogr., **44**, 1272-1276 (2011).